\documentclass[aip,reprint,nofootinbib]{revtex4-1}
\usepackage{graphicx}
\usepackage{bm}

\begin{document}
\title{Cyclotrons with Fast Variable and/or Multiple Energy Extraction }
\date{\today}
\author{C. Baumgarten}
\affiliation{Paul Scherrer Institute, Switzerland}
\email{christian.baumgarten@psi.ch}

\def\begeq{\begin{equation}}
\def\endeq{\end{equation}}
\def\begary{\begeq\begin{array}}
\def\endary{\end{array}\endeq}
\def\bmtx{\left(\begin{array}}
\def\emtx{\end{array}\right)}
\def\eps{\varepsilon}
\def\g{\gamma}
\def\y{\gamma}
\def\w{\omega}
\def\W{\Omega}
\def\s{\sigma}

\def\Exp#1{\exp\left(#1\right)}
\def\Log#1{\ln\left(#1\right)}
\def\Sinh#1{\sinh\left(#1\right)}
\def\Sin#1{\sin\left(#1\right)}
\def\Tanh#1{\tanh\left(#1\right)}
\def\Tan#1{\tan\left(#1\right)}
\def\Cos#1{\cos\left(#1\right)}
\def\Cosh#1{\cosh\left(#1\right)}

\begin{abstract}
We discuss the principle possibility of stripping extraction in combination with reverse bends 
in isochronous separate sector cyclotrons (and/or FFAGs). If one uses reverse bends between 
the sectors (instead of drifts) and places stripper foils at the sector exit edges, 
the stripped beam has a reduced bending radius and it should be able to leave the cyclotron 
within the range of the reverse bend - even if the beam is stripped at less than full energy.

We are especially interested in $H_2^+$-cyclotrons, which allow to double the charge to 
mass ratio by stripping. However the principle could be applied to other ions or ionized 
molecules as well. For the production of proton beams by stripping extraction
of an $H_2^+$-beam, we discuss possible designs for three types of machines: 
First a low-energy cyclotron for the simultaneous production of several beams 
at multiple energies - for instance 15 MeV, 30 MeV and 70 MeV - thus allowing to have beam on several 
isotope production targets. In this case it is desired to have a strong energy dependence of
the direction of the extracted beam thus allowing to run multiple target stations simultaneously.
Second we consider a fast variable energy proton machine for cancer therapy that should allow 
extraction (of the complete beam) at all energies in the range of about 70 MeV to about 250 MeV 
into the same beam line. And third, we consider a high intensity high energy machine, where the
main design goals are extraction with low losses, low activation of components and high reliability.
Especially if such a machine is considered for an accelerator driven system (ADS), this extraction  
mechanism has severe advantages: Beam trips by the failure of electrostatic elements could be
avoided and the turn separation could be reduced, thus allowing to operate at lower
main cavity voltages. This would in turn reduce the number of RF-trips. 

The price that has to be paid for these advantages is an increase in size and/or in field strength 
compared to proton machines with standard extraction at the final energy.
\end{abstract}

\maketitle

\section{Introduction}

A major fraction of the practical problems in the operation of cyclotrons are related to beam
extraction: the activation of extraction elements increases the personal dose during maintenance 
work and sometimes requires to shutoff the beam long before the scheduled work. The
electrostatic elements are frequently the cause of beam interruptions due to high voltage trips, 
they require regular maintenance like cleaning and conditioning. 
In order to increase the extraction efficiency, the energy gain per turn must be maximized, 
which requires to run cavity and resonators at the limit of what can be achieved. 
This in turn increases the frequency of cavity trips and amplifier failures.

In this work, we propose to utilize the mechanism of stripping extraction, which is fairly well-established
in many machines worldwide in nearly the complete energy and intensity range that can be achieved by 
cyclotrons~\cite{strip0,strip1,strip2,strip3,strip4,strip5}. The extraction mechanism that we present 
here might help to avoid most extraction problems completely. In the case of variable energy extraction 
as we propose for proton therapy machines, energy degraders and energy selection systems can be omitted,
thus reducing the costs and the facility footprint significantly. Since the required beam intensities 
are typically in the order of $1\,\mathrm{nA}$ at the patient only, it should be possible to keep
a cyclotron with variable energy extraction almost free from activation of components.
Higher beam currents -- of up to $1\,\mathrm{\mu A}$ -- are mainly required to compensate the
losses of energy degradation and collimation~\cite{G2}.

Variable energy extraction by stripping has been proposed and used at the Manitoba cyclotron by 
stripping of $H^-$-ions~\cite{hminusvarenergystripper} and in the RACCAM cyclotron~\cite{Hminus}. 
Unfortunately, the $H^-$-ion is not stable in strong magnetic fields at high energy, so that 
$H^-$-cyclotron are either limited in energy or restricted magnetic field values. 
This requires large radius machines like the TRIUMF cyclotron~\cite{CraddockSymon}. 
Furthermore, the use of $H^-$ ions in accelerators is more demanding with respect to the machine 
vacuum and the production of $H^-$ in ion sources.

Cyclotrons (and/or FFAGs) with reverse bends have been proposed in the past~\cite{ffag,revbend0,revbend1} -- 
mainly in order to achieve the focusing conditions that are required for energies of $1\,\mathrm{GeV}$
and above. However there is no publication known to the authors that proposes the use of reverse bends in 
combination with stripping extraction. 

In most (if not all) cases where stripping extraction of $H_2^+$-ions is used, the proposed extraction 
schemes lead to complicated orbits that circle one or even multiple times within the cyclotron before 
the beam exits~\cite{strip0,strip1,strip2,strip3,strip4,strip5}. The use of this method for multiple 
or even for continously variable energy extraction is difficult - if at all possible. 

Another method to achieve beam extraction at variable energy is the variation of the main field of 
the cyclotron and to use a sequence of trim coils to achieve isochronism for the desired extraction 
energy. This method is the most ``natural'' way and it is known to work. However the minimal time 
to switch between energies is given by the ramping of the main field and the magnetic relaxation time 
of the yoke. In the optimal case it might be possible to realize energy switching within minutes. 
We are aiming for a millisecond range, i.e. energy switching times that are compatable to the time 
that is required to adjust a beamline with laminated magnets to the new energy.

The goal of this work is to present first basic geometrical and beam dynamical studies in order
to investigate the feasibility of variable energy cyclotrons and to explore the energy ranges that 
could be achieved. Concerning the beam dynamics we restrict ourselves to the minimum, which we
consider to be the verification of the stability of motion of a coasting beam and of the extraction
mechanism. In order to survey the parameter space for such machines we restrict ourselves to the 
so-called hard edge approximation of the magnets. 
And we further simplify this approach by assuming homogeneous magnetic fields within sectors and 
valleys. Isochronism is achieved exclusively by a variation of the azimuthal sector width along 
the orbit~\cite{schatz}.

In Sec.~\ref{sec_geom}, we give a description of the geometry and the calculation of the transfer
matrices. The equations given there have been used in Mathematica\textsuperscript{\textregistered}
to analyze the orbits and the traces of the transfer matrices in hard edge approximation in order 
to find stable solutions with the desired extraction orbits. 
Base on the results, a ``C''-program was used to generate smoothed magnetic field maps, which have 
then been analyzed with an equilibrium orbit code~\cite{Gordon} and a cartesic tracking code to 
verify the analytical results of the hard edge approximation numerically. 

In Sections \ref{sec_isocyc}-\ref{sec_adscyc} we present the results of the calculations.

\section{Geometry of a Separate Sector Cyclotron with Reverse Bends}
\label{sec_geom}

\begin{figure*}
\includegraphics[width=15.5cm]{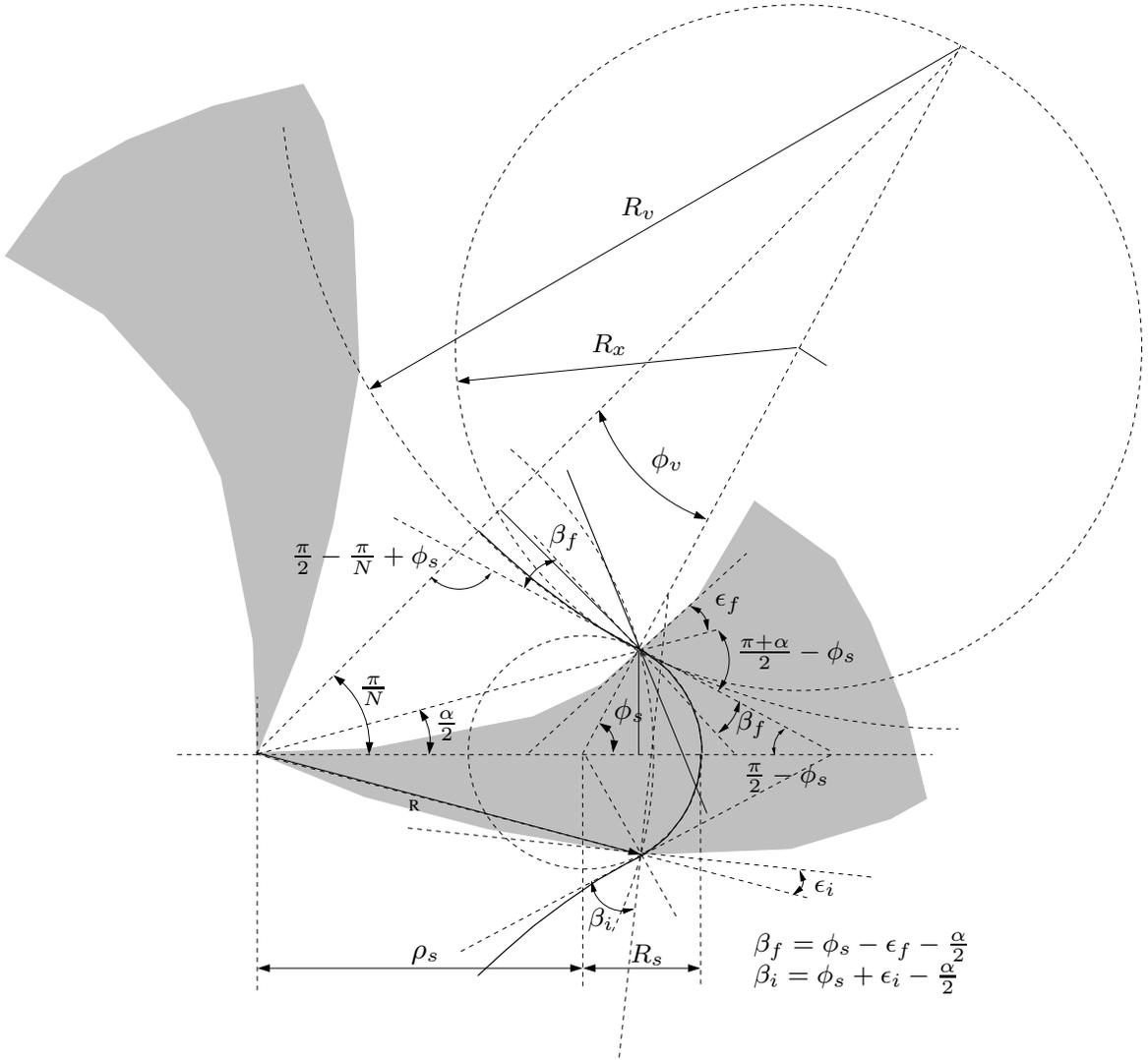}
\caption{
Geometry of a cyclotron ``section'' with reverse bends. The spiralled ``sector'' is indicated by a gray 
polygon. Since it has a constant field, the equilibrium orbit for a certain energy is composed of
two arcs: Within the sector magnet it has bending radius $R_s$ and within the reverse bend (``valley''),
it has a larger radius $R_v$. Due to the increasing complexity of a graphical analysis of the geometry 
we present an Ansatz for an algorithmic method in the App.~\ref{sec_algeom}.
\label{fig_h2cyc4}}
\end{figure*}

We consider $H_2^+$-cyclotrons that are composed of $N$ identical sections which are each composed
of a sequence of homogeneous sector magnets, reverse bends with homogeneous fields and (optionally)
drifts. We do not consider beam injection nor other details of central regions in detail. We are
not concerned about the question, if these machines might need pre-accelerators or can be made 
``compact''. We first consider machines that are composed of exclusively positive and negative
bends as shown in Fig.~\ref{fig_h2cyc4}, where we call the positive bends ``sector'' and the 
negative bends ``valley''.

The absolute values of the sector (valley) field is $B_s$ ($B_v$), the corresponding bending
angle is $2\,\phi_s$ ($2\,\phi_v$), then we have for an ion of mass $m$, charge $q$ and momentum $p=m\,c\,\g\,\beta$:
\begary{rcl}
{\pi\over N}&=&\phi_s-\phi_v\\
R_s&=&{p\over q\,B_s}={m\,c\,\g\,\beta\over q\,B_s}\\
R_v&=&{p\over q\,B_v}={m\,c\,\g\,\beta\over q\,B_v}\\
L_{tot}&=&2\,N\,(R_s\,\phi_s+R_v\,\phi_v)=2\,N\,{m\,c\over q}\,(\frac{\phi_s}{B_s}+\frac{\phi_v}{B_v})\,\beta\,\g\,.
\label{eq_cyc1}
\endary
Isochronism requires that the velocity $v$, the orbital angular frequency $\w_o={2\,\pi\over T}$
and the total length of the orbit $L_{tot}$ are related by
\begary{rcl}
v&=&{L_{tot}\over T}={\w_o\,L_{tot}\over 2\,\pi}\\
\beta&=&{v\over c}={\w_o\,L_{tot}\over 2\,\pi\,c}={L_{tot}\over 2\,\pi\,a}\,,
\label{eq_cyc2}
\endary
where $a={c\over\w_o}$ is the cyclotron length unit. In combination with Eqn.~(\ref{eq_cyc1}) this
yields:
\begary{rcl}
2\,\pi\,a\,\beta&=&2\,N\,{m\,c\over q}\,(\frac{\phi_s}{B_s}+\frac{\phi_v}{B_v})\,\beta\,\g\\
a&=&{N\over\pi}\,{m\,c\over q}\,(\frac{\phi_s}{B_s}+\frac{\phi_s-\pi/N}{B_v})\,\g\\
\phi_s&=&{\pi\over N\,(1+\lambda)}\,\left({B_v\over B_0\,\g}+1\right)\,,
\label{eq_cyc3}
\endary
where we used $\lambda={B_v\over B_s}={R_s\over R_v}$ and defined the ``nominal field'' $B_0$ by
\begeq
B_0={ m\,c\over a\,q}={m\over q}\,\w_o\,.
\endeq
From Fig.~(\ref{fig_h2cyc4}) we pick the following equations
\begary{rcl}
R_v\,\sin{(\phi_v)}&=&R\,\sin{(\frac{\pi}{N}-\frac{\alpha}{2})}\\
&=&R\,\left(\sin{(\frac{\pi}{N})}\,\cos{(\frac{\alpha}{2})}-\cos{(\frac{\pi}{N})}\,\sin{(\frac{\alpha}{2})}\right)\\
R_s\,\sin{(\phi_s)}&=&R\,\sin{(\frac{\alpha}{2})}\,,
\endary
from which we obtain in a few steps
\begeq
\tan{(\frac{\alpha}{2})}={\lambda\,\tan{(\frac{\pi}{N})}\,\tan{(\phi_s)}\over (1+\lambda)\,\tan{(\phi_s)}-\tan{(\frac{\pi}{N})}}\,.
\endeq
If $\theta_c$ is the azimuthal angle of the sector center and $\theta_i$ ($\theta_f$) are the angles
of entrance (exit) of the orbit into the sector, then
\begary{rcl}
\theta_i&=&\theta_c-\frac{\alpha}{2}\\
\theta_f&=&\theta_c+\frac{\alpha}{2}\\
\endary
The angles $\varepsilon_i$ ($\varepsilon_f$) between the sector edges and the radial direction can be
obtained by
\begary{rcl}
\tan{(\varepsilon_i)}&=&R\,{d\theta_i\over dR}=R\,{d\theta_i\over d\g}/{dR\over d\g}\\
\tan{(\varepsilon_f)}&=&R\,{d\theta_f\over dR}=R\,{d\theta_i\over d\g}/{dR\over d\g}\\
\endary
where $R$ is the radius of the orbit entering (exiting) the sector. The angles between the 
orbit and the sector edges (which are required for the transfer matrices) can be computed by
\begary{rcl}
\beta_i&=&\phi_s+\varepsilon_i-\frac{\alpha}{2}\\
\beta_f&=&\phi_s-\varepsilon_f-\frac{\alpha}{2}\,.
\endary
The radii of the arc centers in the sector $\rho_s$ and the valley $\rho_v$ are:
\begary{rcl}
\rho_s&=&R_s\,({\sin{\phi_s}\over\tan{(\frac{\alpha}{2})}}-\cos{\phi_s})\\
\rho_v&=&R_v\,({\sin{\phi_v}\over\tan{(\frac{\pi}{N}-\frac{\alpha}{2})}}+\cos{\phi_v})\\
\endary
The cartesic coordinates $x_s(\phi), y_s(\phi)$ of the orbit inside the sector in dependence of the 
angle $\phi$ can be written as
\begary{rcl}
x_s(\phi)&=&(\rho_s+R_s\,\cos{(\phi)})\,\cos{(\theta_c)}-R_s\,\sin{(\phi)}\,\sin{(\theta_c)}\\
y_s(\phi)&=&(\rho_s+R_s\,\cos{(\phi)})\,\sin{(\theta_c)}+R_s\,\sin{(\phi)}\,\cos{(\theta_c)}\,,
\endary
where $\phi$ ranges from $-\phi_s$ to $\phi_s$. Correspondingly one finds for the valley:
\begary{rcl}
x_v(\phi)&=&(\rho_v-R_v\,\cos{(\phi)})\,\cos{(\theta_c)}-R_v\,\sin{(\phi)}\,\sin{(\theta_c)}\\
y_v(\phi)&=&(\rho_v-R_v\,\cos{(\phi)})\,\sin{(\theta_c)}+R_v\,\sin{(\phi)}\,\cos{(\theta_c)}\,,
\endary
where $\phi$ ranges from $-\phi_v$ to $\phi_v$. 

If we assume that a stripper foil is placed exactly at the sector exit (i.e. at radius $R$ and azimuthal
angle $\theta_f$), then the center coordinates  $x_c(\phi), y_c(\phi)$ of the arc described by the 
extracted orbit are
\begary{rcl}
x_c&=&(\rho_s+(R_s+R_v/2)\,\cos{(\phi_s)})\,\cos{(\theta_c)}\\
&-&(R_s+R_v/2)\,\sin{(\phi_s)}\,\sin{(\theta_c)}\\
y_c&=&(\rho_s+(R_s+R_v/2)\,\cos{(\phi_s)})\,\sin{(\theta_c)}\\
&+&(R_s+R_v/2)\,\sin{(\phi_s)}\,\cos{(\theta_c)}\\
\endary
coordinates $x_x(\phi), y_x(\phi)$ of the extracted orbit are:
\begary{rcl}
x_x(\phi)&=&x_c+R_v/2\,\cos{(\theta_c+\phi_s+\pi-\phi)}\\
y_x(\phi)&=&y_c+R_v/2\,\sin{(\theta_c+\phi_s+\pi-\phi)}\,,
\endary
where $\phi$ starts at zero. With the above equations, we analyzed the geometry of the orbit and
the extraction for different choices of $B_s$, $B_v$, $B_0$ and $\theta_c(\g)$ as a function
of $\g=1+E/E_0$.

\subsection{The Transfer Matrices}

The horizontal transfer matrices ${\bf M}_{s,v}$ for the sector (valley) are given by
\begeq
{\bf M}_{s,v}=\bmtx{cc}
\cos{(2\,\phi_{s,v})}&R_{s,v}\,\sin{(2\,\phi_{s,v})}\\
-{\sin{(2\,\phi_{s,v})}\over R_{s,v}}&\cos{(2\,\phi_{s,v})}\emtx\,.
\endeq
The horizontal transfer matrix that describes the edge focusing effect is
\begeq
{\bf M}_{i,f}=\bmtx{cc}
1&0\\
{\tan{(\beta_{i,f})}\over R_\mathrm{eff}}&1\emtx\,.
\endeq
where $R_\mathrm{eff}=(R_s^{-1}+R_v^{-1})^{-1}$.
Starting with the entrance into the sector magnet the horizontal transfer matrix ${\bf M}$ for a single section 
is the product
\begeq
{\bf M}_{sec}={\bf M}_v\,{\bf M}_f\,{\bf M}_s\,{\bf M}_i\,.
\endeq
The radial focusing frequency $\nu_r$ can be obtained from the parametrization by the twiss-parameters $\alpha_t$, $\beta_t$ and $\g_t$:
\begeq
{\bf M}_{sec}={\bf 1}\,\cos{(2\,\pi\,\nu_r)}+\bmtx{cc}
\alpha_t & \beta_t\\
-\g_t&-\alpha_t\\
\emtx\,\sin{(2\,\pi\,\nu_r)}\,,
\endeq 
from which one obtains with $\beta_t\,\g_t-\alpha_t^2=1$
\begary{rcl}
\cos{(2\,\pi\,\nu_r)}&=&Tr({\bf M}_{sec})/2\\
\sin{(2\,\pi\,\nu_r)}^2&=&Det({\bf M}_{sec}-{\bf 1}\,Tr({\bf M}_{sec})/2)\\
\endary
The matrices for the vertical motion are
\begary{rcl}
{\bf T}_{s,v}&=&\bmtx{cc}
1&2\,R_{s,v}\,\phi_{s,v}\\
0&1\\
\emtx\\
{\bf T}_{i,f}&=&\bmtx{cc}
1&0\\
-{\tan{(\beta_{i,f})}\over R_\mathrm{eff}}&1\emtx\,.
\endary
so that correspondingly
\begeq
{\bf T}_{sec}={\bf T}_v\,{\bf T}_f\,{\bf T}_s\,{\bf T}_i\,.
\endeq
The motion is stable, if 
\begary{rcl}
\vert Tr(M_{sec})/2\vert&\le& 1\\ 
\vert Tr(T_{sec})/2\vert&\le& 1\,.
\endary
In case of cyclotrons with reverse bends, one has a huge flutter $F={\langle B^2\rangle-\langle B\rangle^2\over \langle B\rangle^2}$
due to the negative field regions. If one uses the dimensionless ratios $\lambda={B_v\over B_s}={R_s\over R_v}$ and $\mu={B_0\over B_s}$,
then the flutter yields
\begeq
F={(1-\mu\,\g)\,(\mu\,\g+\lambda)\over\mu^2\,\g^2}\,,
\endeq
and can approch easily values of above $4$. Therefore care must be taken to not have too strong
focusing, i.e. to avoid the $N/2$-stopband. Due to the huge flutter, there is no need for large
spiral angles. In contrary, the spiral angle must be kept sufficiently small to avoid the stopband.

\section{A multibeam isotope production cyclotron}
\label{sec_isocyc}

\begin{figure}
\includegraphics[width=8.5cm]{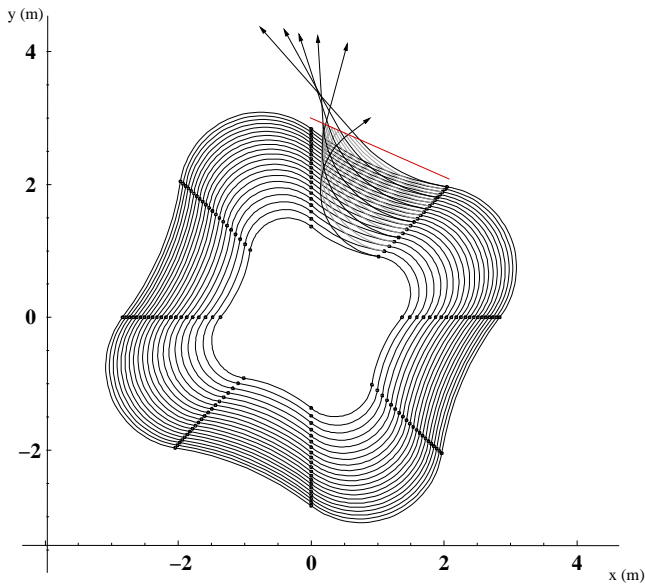}
\caption[]{
Geometry of a $H_2^+$-cyclotron for isotope production with reverse bends. The $H_2^+$-beam is stripped at the sector edge (indicated
by symbols). The orbits of stripped proton beams are shown from $15\,\mathrm{MeV}$ to $70\,\mathrm{MeV}$ in steps of $2.75\,\mathrm{MeV}$.
The arrows indicate the extracted beams for  $15$, $26$, $37$, $48$, $59$ and $70\,\mathrm{MeV}$.
The nominal field $B_0$ is  $0.75\,\mathrm{T}$, the sector field $B_s$ $2\,\mathrm{T}$ and the field strength $B_v$ in 
the reverse bends is $0.55\,\mathrm{T}$.
Without further provision, the directions of the extracted beams differ enough to allow for energy specific targets.
If the beam is partially stripped at lower energy, simultaneous irradiation at several targets should be possible.
If superconducting coils are used to increase the field strength, the size of the cyclotron reduces accordingly.
The sector entrance edge has been chosen straight ${d\theta_i\over d\g}=0$ so that $\theta_c=\theta_i+\alpha/2$ and 
$\theta_f=\theta_i+\alpha$.
\label{fig_h2isocyc}}
\end{figure}

\begin{figure}
\includegraphics[width=8cm]{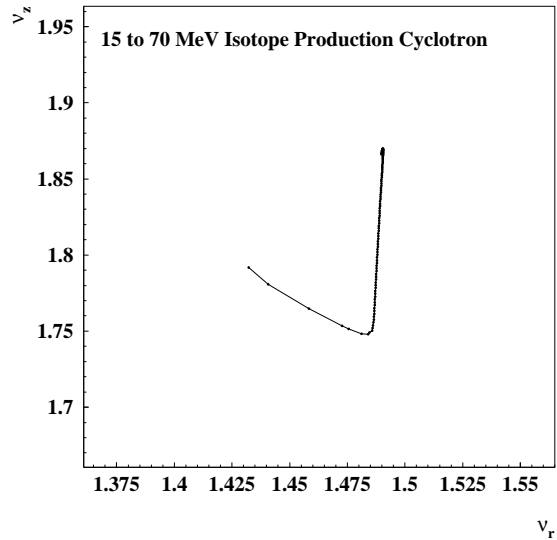}
\caption[]{
Tune diagram of the isotope production cyclotron. The radial tune is (except for the last turns) approximately constant
at about $\nu_r\approx 1.5$. The vertical tune $\nu_z$ starts at low energy at about $1.9$ and decreases smoothly to about
$1.75$ at maximal energy. Due to the large flutter, a stable solution for cyclotron with only 3 sectors has not been found
and we assume that stable solutions can not be found for a comparable energy range of the extracted beam. 
\label{fig_isotunes}}
\end{figure}

\begin{figure}
\includegraphics[width=8cm]{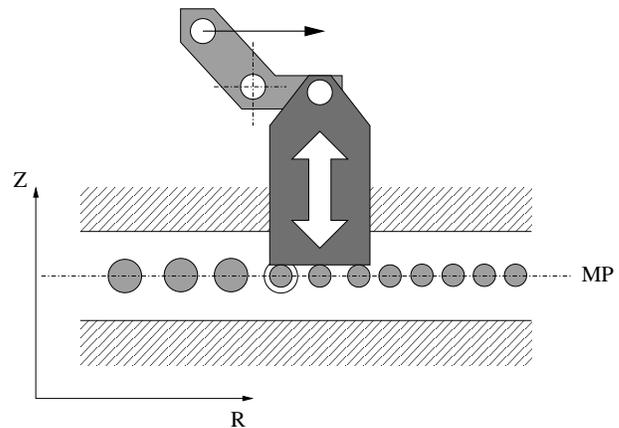}
\caption[]{
Partial beam stripping by vertical positioning of a stripper foil keeping a certain distance to the median plane (MP). 
\label{fig_stripper}}
\end{figure}

The described simplified cyclotron description allows a first analysis, if extraction 
at various energies can be combined with stability of axial and horizontal motion. Fig.~\ref{fig_h2isocyc}
shows a topview layout for a isotope production machine with maximal $H_2^+$-energy of $140\,\mathrm{MeV}$
that allows stripping extraction of proton beams with energies between $15$ and $70\,\mathrm{MeV}$.
Fig.~\ref{fig_isotunes} shows the corresponding tune diagram. Due to the strong flutter, the axial tune $\nu_z$ is very large.
As a consequence the minimum number of sectors is likely $4$, so that the $N/2$-stopband starts at $\nu=2$. Higher sector numbers
are in principle possible, but more expensive and not required for this energy range.
Partial stripping of the beam could allow simultaneous extraction at multiple energies. For this purpose one would move a 
stripper foil vertically towards the median plane until it strips off the desired beam current for the corresponding energy. 
The remaining beam (with reduced emittance) could be accelerated to higher energies
(see Fig.~\ref{fig_stripper}). 

\section{A variable energy cyclotron for proton therapy}
\label{sec_ptcyc}

Commercially available cyclotrons for proton therapy deliver beams with an energy of $235\dots 250\,\mathrm{MeV}$~\cite{Klein, Jongen}.
Since the presently available cyclotron technology delivers the beam at fixed energy, the beam energy must be reduced to
the value that is required for the treatment. This is typically done by energy degradation at the cost of significant emittance 
increase and energy straggling in the degradation process~\cite{Deg0,Deg1,Deg2}. In order to deliver a beam of the required 
quality most of the degraded beam has to be cut off by collimators and an energy selection system (ESS). 
The intensity is (depending on energy) reduced by up to three orders of magnitude. 

Even though there are strong arguments for the use of cyclotrons in proton therapy, there are also
disadvantages of the combination of a fixed-energy-cyclotron, degrader and ESS:
\begin{enumerate}
\item the strong energy dependence of the beam intensity which makes fast and save energy variations (without intensity variations)
of the beam difficult to achieve.
\item the activation of the accelerator, the degrader material, the collimators and other components,
which could be reduced by orders of magnitude, if one could extract high quality beam at various energies.
\item the cost for the degrader and the ESS which typically consists of two dipoles, eight quadrupoles, moveable slits,
beam diagnostics and vacuum components for about $10\,\mathrm{m}$ beamline.
\item the need to use large aperture quadrupoles and dipoles in order to achieve a suitable transmission efficiency
of beam line and gantry.
\end{enumerate}
The list is certainly incomplete, but it suffices to argue that one has to take the over-all costs of an accelerator concept 
into account. A separate sector cyclotron with reverse bends is certainly more expensive than a compact cyclotron. It will also
have a larger footprint and a higher power consumption. However the footprint of the accelerator itsself is only
a small fraction of a complete proton therapy facility. 
\begin{figure}
\includegraphics[width=8.0cm]{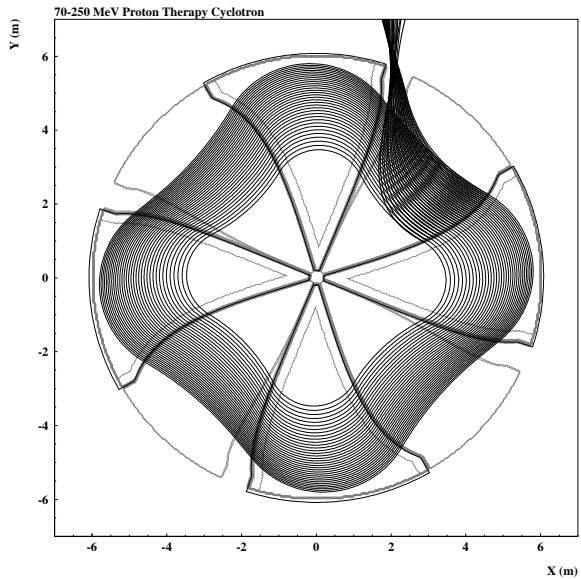}
\caption[]{
Geometry of a $H_2^+$-cyclotron for proton therapy with reverse bends and drifts. The equilibrium orbits and  
stripped proton trajectories for energies from $70\,\mathrm{MeV}$ to $250\,\mathrm{MeV}$ in steps of 
$\approx 5.5\,\mathrm{MeV}$ are also shown. They been computed by an equilibrium orbit code~\cite{Gordon} and by
Runge-Kutta tracking, respectively.
The nominal field $B_0$ is  $0.7\,\mathrm{T}$, the sector field $B_s$ $2\,\mathrm{T}$ and the field strength $B_v$ in 
the reverse bends is $0.55\,\mathrm{T}$. With a bit more fine-shaping of the magnetic field of the reverse bend, it should
be possible to make all extracted beams pass a region small enough to install a fast ``catcher'' magnet, which allows
to bend the extracted beams of all energies into the same beamline. The resulting tunes are shown in the right graph.
The small spiral angle has been introduced to avoid the $\nu_r=\nu_z$-resonance shown as a solid straight line. 
Using superconducting coils and correspondingly higher field values, the size could be reduced accordingly. 
If partial stripping would be applied, it should be possible to extract up to four beams simultaneously.
\label{fig_medcyc}}
\end{figure}

We found that variable energy extraction by a moveable stripper foil before a reverse bend allows {\it in principle}
to extract beams with energies between $70\,\mathrm{MeV}$ and $250\,\mathrm{MeV}$. Fig.~\ref{fig_medcyc} shows the layout 
of an $H_2^+$-cyclotron with $2\,\mathrm{T}$ sector magnets and $0.55\,\mathrm{T}$ reverse bend magnets, the equilibrium
and extraction orbits for energies from $70\,\mathrm{MeV}$ to $250\,\mathrm{MeV}$. The use of superconducting coils
would allow to increase the field and the radius would reduce by the same factor.
\begin{figure}
\includegraphics[width=7.5cm]{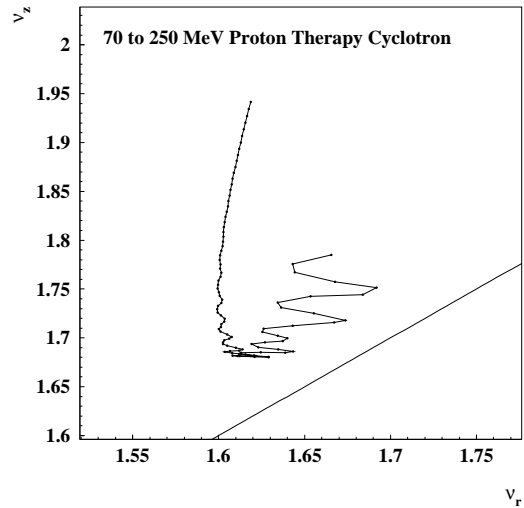}
\caption[]{
Tune diagram of the medical proton cyclotron with variable energy extraction as shown in Fig.~\ref{fig_medcyc}.
\label{fig_medcyctunes}}
\end{figure}

The time required for a change of the beam energy is then determined by the ramping time of the beamline magnets
and the time for the positioning of the stripper foil. If a series of foils at different radii would be inserted 
vertically into the beam, then the actuator would need just a few millimeters of motion for the insertion of the foil
as shown in Fig.~\ref{fig_stripper}. Other mechanisms using radial motion with the advantage of
continous energy adjustment are also possible. Even though the design of fast moveable parts in vacuum is not 
trivial, we believe that mechanisms should be feasible with a response time in the order of $100\,\mathrm{ms}$ or below.
Since the extracted beam current that is required for radiation therapy is of the order of $1\,\mathrm{nA}$, 
cooling of the stripper foil is (for this application) not necessary.

More challenging (in terms of costs and engineering time) is the design of a central region that allows
either to use an internal ion source or a spiral inflector. An internal ion source causes a higher rest gas 
pressure compared to an external source. However the beam current in such a PT machine is very low
so that even a high relative beam loss by rest gas stripping could be accepted.

Certainly the presented extraction mechanism could also be used in combination with a pre-accelerator,
but the stripping process itsself can be used only once. The preaccelerator would necessarily
have a different extraction mechanism.

The design scetched in Fig.~\ref{fig_medcyc} has four sectors so that with an appropriate design of 
rf-resonators one might use at maximum four exit ports in four directions. 
They might (but don't have to) be used simultaneously in order to deliver beam for four treatment rooms 
located around the cyclotron bunker. Since a direct beam from the cyclotron has a small emittance and 
energy spread, the beam transport system does not require magnets with large aperture. Hence beamline
and gantry might be smaller and cheaper than those of conventional systems.
If the beam size and energy spread are too small for fast painting of the tumor, one could insert
scatterers into the beam path - or one might directly use ``thick'' stripper foils, which increase the 
beam size by scattering and make the beam shape more Gaussian.

We used the flat field design since it allows to calculate the desired properties analytically in 
very good approximation. However a cyclotron with a flat field has also practical advantages. It allows for 
instance to make precise online field measurements by NMR-probes. The results could be used to stabilize 
the magnetic field without beam extraction as it is required for a phase probe~\cite{phase}. This would not only
reduce start up time and simplify beam quality management, but it might also reduce activation of an external
beam dump. Furthermore the mechanism that places the stripper foil - if fast enough - could be made ``fail-save'':
if a spring retracts the foil off the median plane in case of emergency, extraction immediately stops.
Without stripper foil but with an appropriate shaping of the edge field with enough phase shift per turn, 
the cyclotron could operate in a stand-by mode without activation and extraction but with contineous beam 
in the median plane. The beam would be accelerated to maximal energy, phase shifted in the fringe field, 
decelerated back to the cyclotron center and dumped there without activation of components. In this way, 
the equipment could stay ``warm'' in stand-by mode. If beam is requested, the only action to be taken is to 
insert the stripper foil at the desired location for the requested energy.

\section{A high energy high intensity proton cyclotron}
\label{sec_adscyc}

Recently there has been renewed interest in high intensity cyclotrons not only for the potential use
in accelerator driven systems (ADS) for transmutation of nuclear waste or as ``energy amplifier''~\cite{EA1}, but
also for physical experiments like $DAE\delta ALUS$~\cite{Daedalus1,Daedalus2,Daedalus3}. 
Typically the cyclotron should be able to deliver $10\,\mathrm{mA}$ or more proton beam current
at $800$ and $1000\,\mathrm{MeV}$. Such cyclotrons have never been build, but the PSI ring machine 
which delivers $2.2\,\mathrm{mA}$ at $590\,\mathrm{MeV}$ often serves as a proof-of-principle machine~\cite{ring}. 
However, there is still a factor of $8$ between the beam power of the PSI machine ($1.3\,\mathrm{MW}$) 
and the desired $10\,\mathrm{MW}$ (or more) for an ADS driver. We are not going to discuss this in 
detail here, but we give an example of an $H_2^+$-cyclotron with stripping extraction between $500$ and 
$950\,\mathrm{MeV}$. The major advantage of the proposed extraction method is the increased reliability 
of extraction without electrostatic elements. Furthermore the concept allows to reduce the turn separation, 
i.e. the required voltage of the accelerating cavities is not a major design issue. A flattop system could
also be obsolete.

\begin{figure}
\includegraphics[width=8.0cm]{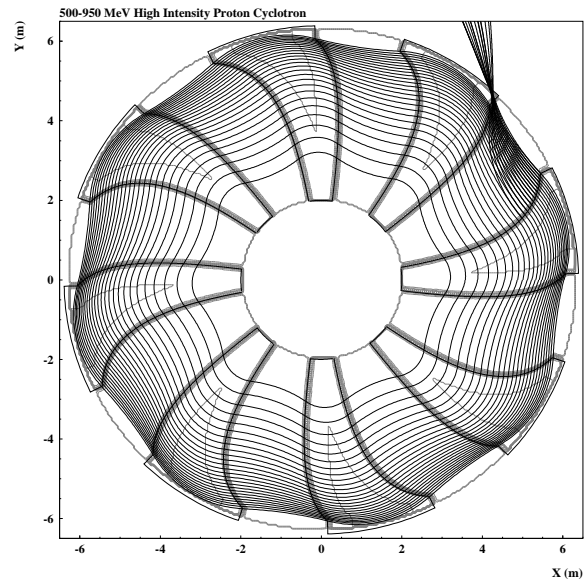}
\caption[]{
Geometry of an 8-sector $H_2^+$-cyclotron for extraction energies in the range between $500$ and $950\,\mathrm{MeV}$ 
for ADS. The orbits of the protons after stripping are shown from $470\,\mathrm{MeV}$ to $950\,\mathrm{MeV}$.
The nominal field $B_0$ is  $0.88\,\mathrm{T}$, the sector field $B_s$ $4\,\mathrm{T}$ and the field strength $B_v$ in 
the reverse bends is $1.05\,\mathrm{T}$. The tune diagram is shown in the right graph and covers the $H_2^+$-energy 
range from $220\,\mathrm{MeV}$ to $1.9\,\mathrm{GeV}$. A more advanced field shaping with reduced flutter at lower
energies would keep the vertical $\nu_z$ below $2\,\nu_r$ also at lower injection energy and would allow to stay 
above $\nu_r$ at higher energies. The spiral angle has been chosen to be $\theta_c=(\g-1)/4.2$, $\theta_i=\theta_c-\alpha/2$
and $\theta_f=\theta_c+\alpha/2$.
\label{fig_adscyc}}
\end{figure}

\begin{figure}
\includegraphics[width=7.5cm]{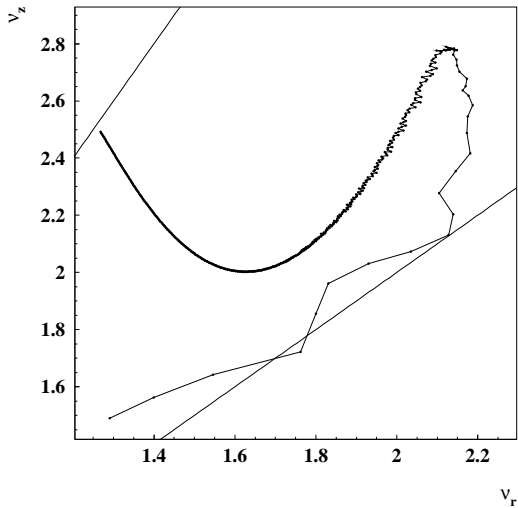}
\caption[]{
Tune diagram of the high intensity cyclotron with variable energy extraction as shown in Fig.~\ref{fig_adscyc}.
\label{fig_adscyctunes}}
\end{figure}

Fig.~\ref{fig_adscyc} shows a machine layout for a homogenous sector (valley) field of $4\,\mathrm{T}$
and $1.05\,\mathrm{T}$. As shown in Fig.~\ref{fig_adscyctunes}, major resonances could be avoided by an
adequat choice of the spiral angle. 
There are two major differences between the cyclotron design here and the one proposed in Ref.~\cite{Daedalus3}, the
first being the difference in the vertical tune, which is in our design considerably increased by the reverse bends. 
The second is the trajectory of the stripped beam. The design proposed in Ref.~\cite{Daedalus3} uses the 
conventional scheme in which the stripped beam is bend inwards and passes the cyclotron median plane at 
nearly all radii before exiting the field. There is no principle problem with this scheme, but it has 
disadvantages. First the exact position and direction of the extracted beam depends on the cyclotron field 
all along the extraction orbit which is more or less the complete median plane area. It is therefore influenced 
by trim coils, cavities and main field changes. Second, this beam path has to be free from obstacles. 
Fig.~\ref{fig_adscyc} shows a machine layout for a homogeneous sector (valley) field of $4\,\mathrm{T}$
and $1.05\,\mathrm{T}$, Fig.~\ref{fig_adscyctunes} the tune-diagram. The spiral angle has been optimized to
avoid major resonances. 
The extraction path with reverse bends is as short as possible and passes only the area between two sectors. 
It is therefore much less sensitive to changes of main field and/or trim coil settings.

\section{Some final remarks}

The discussed machine layouts allow some further optimization with respect to the direction of the extracted
beam by an appropriate shaping of the fields of the sectors magnets and the reverse bends. This would go beyond
the scope of this paper, since our intention was to survey the {\it principle} possiblities of the extraction 
mechanism. Certainly the flat field approach used above is neither necessary for this extraction scheme to work 
nor do we consider it to be the optimal choice. It has been chosen as it allows for a fairly simple analytical 
description of cyclotron beam optics.

The ``inner region'' of such cyclotrons, i.e. the energy range in which beam extraction is not possible, 
might be designed very different from what is scetched above. The negative field in the reverse bends is 
not required at small radii. Therefore it is possible (and unavoidable) to reduce the effect of the reverse 
bend towards the cyclotron center (compensating this with reduced sector field or sector
width). 

The beam in a cyclotron like the ones described above should be centered so that the energy and radius are related 
in a predictable and reproducable way. This is especially important for PT applications, where the position 
of the stripper foil selects the beam energy.

Since resonant beam extraction is not required, the phase curve may be chosen flat up to the cyclotron fringe 
field. This allows to accelerate beams with relatively low cavity voltages. 

Since neither a low energy spread nor a high turn separation is essential in order to minimize extraction losses 
(depending on the acceptance of the beam line transporting the beam to target), even the high intensity machine 
might be operational without flattop cavity. The space saved this way could be used to improve the vacuum
conditions by the installation of cryogenic pumps. The beam loss by rest gas stripping has to be minimized
when such cyclotrons are operated with high currents. 

We discussed a long list of advantages of the new extraction mechanism, but the discussed method has 
it's price: $H_2^+$ has half the charge to mass ratio of protons and therefore on has to use double
size and/or field strength to reach the same final proton energy. The use of reverse bends has a comparable
effect. A discussion, if and when the increase in size or field strength pays off by the mentioned advantages, 
is beyond the scope of this paper, but depends certainly on the purpose of the machine. 

\section{Summary}

The geometry of cyclotrons with reverse bends has been analyzed and the resulting transfer matrices have been 
given. We investigated some of the design options involving the use of reverse bends in combination with stripping 
extraction of $H_2^+$. We proved the principle feasability of variable energy/multiple beam extraction from 
cyclotrons with reverse bends and verified the analytical beam stability by a numerical calculation of the tunes. 

We presented three potential applications for the described extraction mechanism, 
an isotope production cyclotron with simultaneous extraction at several energies between $15$ and $70\,\mathrm{MeV}$, 
a medical cyclotron with a variable energy extraction in the range between $70$ and $250\,\mathrm{MeV}$ and a 
high intensity ring cyclotron with beam extraction at energies between $500$ and $950\,\mathrm{MeV}$.

\section{Acknowledgements}

We thank Nada Fakhoury for her help in writing the Mathematica\textsuperscript{\textregistered} notebooks used for this work.

Software has been written in ``C'' and been compiled with the GNU\textsuperscript{\copyright}-C++
compiler on Scientific Linux. The figures have been generated with the cern library (PAW) and XFig.

\begin{appendix}
\section{An algebraic method for the analysis of accelerator floor layouts}
\label{sec_algeom}

The floor layout of cyclotrons is just a special case of the general problem
of the calculation of floor layouts, which itself is a special case of the geometry
of curves in the plane. In the general case, a planar smooth curve can be described
by a ``state vector'' $\psi$ that contains the coordinates and the direction
derivatives ${\bf \psi}=(x,y,x',y')$, where $x$ and $y$ are the Cartesic
coordinates of the orbit (planar curve) and $x'={dx\over ds}$ and $y'={dy\over
  ds}$ are the derivatives with respect to the pathlength $s$. 
By definition one has
\begeq
x'^2+y'^2=1\,,
\endeq
so that one may also write $(x',y')=(\cos{\phi},\sin{\phi})$ with the direction
angle $\phi$ of the orbit.

The state vector is a function of the pathlength $s$ of the orbit and the general
evolution of this vector can be described by a differential equation of the
form:
\begary{rcl}
{\bf\psi'}&=&{d{\bf\psi}\over ds}={\bf F}({1\over\rho})\,{\bf\psi}\\
\bmtx{c}
x\\y\\x'\\y'\\
\emtx&=&
\bmtx{cccc}
0&0&1&0\\
0&0&0&1\\
0&0&0&-\frac{1}{\rho}\\
0&0&\frac{1}{\rho}&0\\
\emtx\,
\bmtx{c}
x\\y\\x'\\y'\\
\emtx\,,
\endary
where $\rho=\rho(s)$ is the local bending radius of the curve.
In the hard edge approximation, we assume that $\frac{1}{\rho}={q\,B\over
  p}$ is piecewise constant. In this case, a transfer matrix method can be
used and the solution is given by a transfer (or transport) matrix ${\bf
  M}(s)$:
\begeq
{\bf\psi}(s)={\bf M}(s)\,{\bf\psi}(0)\,,
\label{eq_solution}
\endeq
where the matrix ${\bf M}$ is the product of the transfer matrices for the
individual segments:
\begeq
{\bf M}(s)=\prod\limits_{k=0}^{n-1}\,{\bf M}_k
\endeq
In hard edge approximation, there are basically two transfer matrices, the
matrix ${\bf M}_d(L)$ for a drift of length $L$ and the matrix ${\bf
  M}_b(\rho,\alpha)$ 
for a bending magnet for a bending radius $\rho$ and angle $\alpha$:
\begary{rcl}
{\bf M}_d(L)&=&\exp{({\bf F}(0)\,L)}\\
{\bf M}_b(\rho,\alpha)&=&\exp{({\bf F}({1\over\rho})\,\alpha\,\rho)}\\
\endary
The matrix powers of ${\bf F}$ are readily computed:
\begary{rcl}
{\bf F}^2&=&\bmtx{cccc}
0&0&0&-{1\over\rho}\\
0&0&{1\over\rho}&0\\
0&0&-{1\over\rho^2}&0\\
0&0&0&-{1\over\rho^2}\\
\emtx\\
{\bf F}^3&=&-{1\over\rho^2}\,{\bf F}\\
{\bf F}^4&=&-{1\over\rho^2}\,{\bf F}^2\\
\endary
from which one finds within a few steps
\begary{rcl}
{\bf M}_b(\rho,\alpha)&=&\bmtx{cccc}
1&0&\rho\,s&-\rho\,(1-c)\\
0&1&\rho\,(1-c)&\rho\,s\\
0&0&c&-s\\
0&0&s&c\\
\emtx\,,
\endary
where $s=\sin{\alpha}$, $c=\cos{\alpha}$ and $\alpha={L\over\rho}$. 
A reverse bend (i.e. a bend into the opposite direction), is described
by a negative radius {\it and} a negative angle, yielding a positve length $L=\alpha\,\rho$.
If ${1\over\rho}=0$, then the transfer matrix simplifies to the transfer matrix of a drift:
\begary{rcl}
{\bf M}_d(L)&=&\exp{({\bf F}\,L)}=\bmtx{cccc}
1&0&L&0\\
0&1&0&L\\
0&0&1&0\\
0&0&0&1\\
\emtx\,.
\endary
These two matrices are sufficient to compute the floor layout of most
accelerator beamlines. But they are also usefull for the geometrical
analysis of separate sector cyclotrons with homogeneous field magnets
in hard edge approximation as described above. 

In addition to the above transfer matrices, we will use the familiar
coordinate rotation matrix ${\bf M}_{rot}(\theta)$:
\begeq
{\bf M}_{rot}(\theta)=\bmtx{cccc}
\cos{\theta}&-\sin{\theta}&0&0\\
\sin{\theta}&\cos{\theta}&0&0\\
0&0&\cos{\theta}&-\sin{\theta}\\
0&0&\sin{\theta}&\cos{\theta}\\
\emtx\,.
\endeq
If we consider an accelerator with $N$ equal sectors (or sections), then - given
an arbitrary starting position ${\bf\psi}(0)$ - the position and direction change 
after one sector {\it relative to some center point ${\bf\psi}_c$} can be described 
by a rotation with an angle of $\theta={2\,\pi\over N}$. Hence we may write
\begeq
({\bf\psi}(L)-{\bf\psi}_c)={\bf M}_{rot}(\theta)\,({\bf\psi}(0)-{\bf\psi}_c)\,,
\endeq
so that by the use of Eqn.~\ref{eq_solution} one finds
\begary{rcl}
({\bf M}_{sec}{\bf\psi}(0)-{\bf\psi}_c)&=&{\bf M}_{rot}(\theta)\,({\bf\psi}(0)-{\bf\psi}_c)\\
({\bf M}_{sec}-{\bf M}_{rot}(\theta))\,{\bf\psi}(0)&=&({\bf 1}-{\bf M}_{rot}(\theta))\,{\bf\psi}_c\,.
\endary
The coordinates of the accelerator center can be obtained by:
\begeq
{\bf\psi}_c=({\bf 1}-{\bf M}_{rot}(\theta))^{-1}\,({\bf M}_{sec}-{\bf M}_{rot}(\theta))\,{\bf\psi}(0)\,.
\label{eq_center}
\endeq
The matrix ${\bf M}_x(\theta)\equiv ({\bf 1}-{\bf M}_{rot}(\theta))^{-1}$ can be directly computed
and is explicitely given by
\begary{rcl}
{\bf M}_x(\theta)&=&{1\over 2\,\sin{({\theta/2})}}\,{\bf M}_{rot}(\pi/2-\theta/2)\\
&=&\frac{1}{2}\,\bmtx{cccc}
1&-\cot{\theta/2}&0&0\\
\cot{\theta/2}&1&0&0\\
0&0&1&-\cot{\theta/2}\\
0&0&\cot{\theta/2}&1\\
\emtx
\endary
If one computes the center of motion of a bending magnet for an angle $\theta$, the result is given by
\begary{rcl}
{\bf\psi}_c&=&{\bf M}_x(\theta)\,({\bf M}_b(\rho,\theta)-{\bf M}_{rot}(\theta))\,{\bf\psi}(0)\\
&=&\bmtx{cccc}
1&0&0&-\rho\\
0&1&\rho&0\\
0&0&0&0\\
0&0&0&0\\
\emtx\,{\bf\psi}(0)\\
&=&(x(0)-\rho\,y'(0),y(0)+\rho\,x'(0),0,0)^T\,,
\label{eq_centerbend}
\endary
which is easy to verify.

The computation of the center coordinates is therefore straightforward - and yields
a result even, if the matrix ${\bf M}_{sec}$ does not describe a ``valid'' sector. 
Such a non-valid situation is given, if the  the ``velocity'' components of ${\bf\psi}_0$ 
do not vanish, which happens, if the sum of the bending angles entering ${\bf M}_{sec}$ does
not equal $\theta$. 

In the following we use Eqn.~\ref{eq_center} to compute the starting conditions for a
cyclotron centered at $x_c=y_c=0$, i.e. the radius and direction of an {\it equilibrium orbit}.
If we let the orbit start at $(0,0)$ in arbitrary direction, i.e. we choose 
for instance ${\bf\psi}(0)=(0,0,0,1)$, then the orbit with starting position ${\bf\psi}(0)-{\bf\psi}_c$ 
is centered. Hence the starting position
\begeq
{\bf\psi}(0)=\left({\bf 1}-{\bf M}_x(\theta)\,({\bf M}_{sec}-{\bf M}_{rot}(\theta))\right)\,(0,0,0,1)^T\,.
\label{eq_centerstart}
\endeq
is centered. The orbit still starts at an ``arbitrary'' angle $\theta_0$, i.e. ${\bf\psi}(0)$ as 
given by Eqn.~\ref{eq_centerstart} can be written as
\begeq
{\bf\psi}(0)=(R\,\cos{\theta_0},R\,\sin{\theta_0},0,1)^T=(x_0,y_0,0,1)^T\,.
\endeq
If one aims for a specific orientation of the orbit with respect to the floor coordinates - for instance 
on the x-axis - then one may use the rotation matrix with $\theta_0=\arctan{({y_0\over x_0})}$:
\begeq
{\bf\psi}(0)\to {\bf M}_{rot}(-\theta_0)\,{\bf\psi}(0)=(R,0,\frac{y_0}{R},\frac{x_0}{R})^T\,. 
\endeq
The angular width of the magnet can then be calculated by computing the position angle of ${\bf M}_b\,{\bf \psi}(0)$.

\begin{figure}
\includegraphics[width=8.5cm]{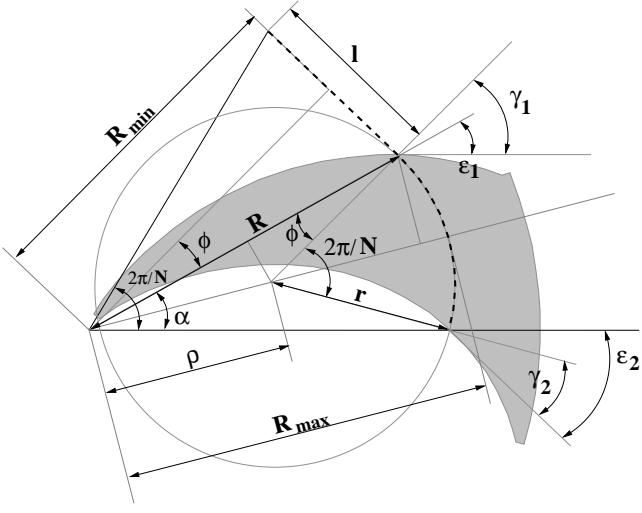}
\caption[]{
Geometry of the cyclotron sector in case of a sector magnet (gray area) with a constant field 
along the closed orbit (shown in as a thick dashed line). 
$\eps_1$ and $\eps_2$ are the spiral angles of the entrance and exit of the magnet.
$\g_1$ and $\g_2$ are the angles between the sector entrance and exit and the orbit normal vector.
It is obvious from the drawing that $R\,\sin{(\alpha/2)}=r\,\sin{(\pi/N)}$.
\label{fig_secgeom}}
\end{figure}
If this method is applied to a cyclotron sector composed of a dipole with bending radius $r$ and 
bend angle ${2\,\pi\over N}$ and a drift of length $L$ as shown in Fig.~\ref{fig_secgeom}, then one 
obtains:
\begary{rcl}
\tan{\phi}&=&{2\,r\over L}+\cot{\pi\over N}\\
R&=&\sqrt{\frac{L^2}{4}+(r+\frac{L}{2}\,\cot{\pi\over N})^2}\,.
\endary
Both conditions could also be derived from Fig.~\ref{fig_secgeom}. The
advantage of the algebraic method is, that it gives an {\it algorithm} 
at hand that allows to determine the essential geometric conditions directly
from the parameters $r$, $N$ and $L$, without the need to analyze a ``hand-made'' drawing. 
Furthermore the algebraic algorithm enables to produce the drawing. 

In case of the simple situation scetched in Fig.~\ref{fig_secgeom}, the drawing might do as well.
But in the case of the medical cyclotron as shown in Fig.~\ref{fig_medcyc}, the symmetrie
of the equilibrium orbit for a given energy is broken by the drift between the reverse
bend (valley) and the next sector. In this case and in case of more complex configurations, 
the analysis of the layout by a handmade scetch becomes cumbersome due to the increasing number 
of angles and geometrical relations. In fact, the geometry of the medical cyclotron 
presented above has been analyzed by a ``C''-program and a Mathematica\textsuperscript{\textregistered} 
notebook based on the above algebraic ansatz. The main reason was the desire to create a map of 
the magnetic field in cylindrical coordinates for the numerical (and hence more ``realistic'') 
computation of the tunes.
In case of a cyclotron that is composed of $N$ sections each containing a sector magnet, a reverse
bend and a drift, it turns out, that the entrance and exit radius for a given energy are not equal.

For a given radius of the grid, we had to determine the energy (i.e. the $\g$-value) of the
equilibrium orbit entering the sector, a second $\g$-value for the orbit existing the sector and
a third one at the exit of the reverse bend. This was done by an iterative numerical interval search.

\end{appendix}

\end{document}